\documentclass[letterpaper,10pt]{article}

\title{On accretion of dark energy onto black- and worm-holes}
\author{Jos\'e A. Jim\'enez Madrid$^{1,}$\footnote{Electronic
    adresses:J.Madrid@damtp.cam.ac.uk, madrid@imaff.cfmac.csic.es}\  and
  Prado Mart\'{\i}n-Moruno$^{2,}$\footnote{Electronic address: pra@imaff.cfmac.csic.es} \\
{\small $^{1}$Department of Applied Mathematics and Theoretical Physics,} \\
{\small Wilberforce Road, Cambridge, CB3 0WA,
United Kingdom} \\
{\small $^{2}$Colina de los Chopos, Instituto de F\'{\i}sica Fundamental,}\\
{\small Consejo Superior de Investigaciones Cient\'{\i}ficas, Serrano 121, 28006 Madrid, Spain}}

\begin{document}

\maketitle

\begin{abstract}
We review some of the possible models that are able to describe the
current Universe which point out the future singularities that could
appear. We show that the study of the dark energy accretion onto
black- and worm-holes 
phenomena in these models could lead to unexpected consequences,
allowing even the avoidance of the considered singularities. We also
review the debate about the approach used to study the accretion
phenomenon which has appeared in 
literature to demonstrate the advantages and drawbacks of the different points of view. We finally suggest new lines of research to resolve the shortcomings of the different accretion methods. We then discuss future directions for new possible observations that could help choose the most accurate model.
\end{abstract}

\section{Introduction}

The discovery of the cosmic acceleration indicated by the
observational data \cite{Mortlock:2000zu,Riess:1998cb,Spergel:2003cb}
has caused a break in the belief of what could be the matter content
of the universe and what might be its possible future evolution. The
interpretation of this data in the framework of General Relativity
implies that the majority of the content in the universe should be 
new stuff, which has been called dark energy, possessing
anti-gravitational properties, i.e. the equation of state parameter of
the dark energy must be $w<-1/3$ ($w=p/\rho$). It even seems to be possible
that the equation of state parameter is less than $-1$. In
that case the new stuff is known as phantom energy \cite{Caldwell:1999ew} and the consideration of this fluid
could lead the universe to a catastrophic end by the appearance of a
future singularity. The most popular of such singularities is the
so-called big rip \cite{Caldwell:2003vq}, which is a possible doomsday
of the universe where both of its size and its energy density become
infinitely larger. However, the big rip is not the only possibility
which has been suggested as the catastrophic end to the universe in the
new phantom models. It has also been argued that the universe could
finish its evolution at a time where its energy density becomes
infinitely larger maintaining the scale factor as a finite value, known as
the big 
freeze singularity \cite{BouhmadiLopez:2006fu,BouhmadiLopez:2007qb}.

It is well known that dark energy should be accreted onto black holes
in a different way that ordinary matter does, since that new fluid
covers the whole space. Therefore, the study of dark energy accretion
onto black holes becomes an interesting field of study which could
lead to surprising effects as the possible disappearance of black
holes in phantom environments. As we will show in this chapter the
mentioned accretion phenomenon was originally studied by Babichev et
al. \cite{Babichev:2005py}, although a great number of works have
been done to improve the method used by those 
authors \cite{MartinMoruno:2006mi,Gao:2008jv,MartinMoruno:2008vy}.

On the other hand,  the accretion process could also imply unexpected consequences in the case that one considers the evolution of cosmological objects even stranger than black holes, wormholes\footnote{For information about the historical development of wormholes and a deep study of their spacetime, please see Ref.\cite{Visser}}.
Traversable wormholes are short-cuts between
two regions of the same universe or between two universes, which could be
used to construct time-machines \cite{Morris:1988tu}. The reason for its strangeness is not related to
its bridge character but to that, in order to be traversable and stable, the wormholes
must be surrounded by some kind of exotic matter which do not fulfil the null
energy condition. Nevertheless, the consideration of phantom models as
the possible current description of the universe has caused a 
revival of interest in traversable wormholes, since this fluid would also violate the null energy condition. Even more, it has been shown that an inhomogeneous
version of phantom energy can be the exotic stuff which supports
wormholes \cite{Sushkov:2005kj}. In some cosmological
models the accretion of phantom energy onto a wormhole could lead to
an enormous growth of its mouth, engulfing the whole universe which
would travel through it in a big trip \cite{GonzalezDiaz:2005yj},
avoiding the big rip \cite{GonzalezDiaz:2004vv} or big 
freeze \cite{BouhmadiLopez:2006fu} singularity in the corresponding cases.

In the present chapter we show the method of how to treat the
accretion phenomenon onto black- and worm- holes and its 
cosmological consequences in some models. In Sec.~II, we review
some candidates that could be responsible for the current 
cosmological acceleration and which pay special attention to the possible
future singularities appearing in some of them. 
In Sec.~III, the procedure of the study of the dark energy accretion
onto black holes
based on the Babichev et al. method is shown and 
its application to the models included in Sec.~II is presented. The
corresponding study in the case of accretion onto 
wormholes is shown in Sec.~IV, where the possible avoidance
of the future singularities is highlighted. Since the study of 
dark energy accretion onto black-
and worm- holes is still an open issue, we refer 
in Sec. V some interesting works which have produced a debate
about the used method and we also outline 
some possible lines for future research to solve the
shortcomings. Finally, in Sec.~VI, the results are discussed and further comments are added.

\section{Brief review of some candidates to cosmic acceleration.}
The origin of the current accelerating expansion of the universe is one of the most interesting challenges in cosmology. Therefore, a plethora of cosmological models have been developed in recent years in order to take into account such acceleration.
Although modifications of the Lagrangian of General Relativity or
considerations of more than four dimensions could explain the current
phase of the universe, the acceleration can also be modelled in the
framework of General Relativity theory, which has shown 
agreement with the observational tests up to now. Nevertheless, the
consequences of using the Einstein's theory is 
that most part of the universe's content must 
be some kind of fluid with anti-gravitational properties, called dark energy.

In order to show the necessity of the inclusion of dark energy as a
new component of a universe described by General Relativity, we must
consider an homogeneous, isotropic and spatially flat universe, i.e. a
Friedmann-Lema\^{\i}tre-Robertson-Walker (FLRW) model with $k=0$. As an
approximation, one can consider that this model is only 
filled with one fluid\footnote{It must be noted that such an
  approximation is justified because the contribution of 
the current dark energy density is around $74\%$ of the total energy
density of the universe and, from the evolution of the energy density
in terms of the scale factor, it is expected that dark energy density decays slower than the ordinary matter energy density when the scale factor increases, being therefore the future dynamic of the universe governed by the dark component. Even more, in the phantom case the phantom energy density would increase with the scale factor, so the same conclusions can be, of course, recovered in this case.}.
Throughout this chapter we shall use natural units so that G = c = 1.
The Friedmann equations can be expressed in the usual way \cite{Wald}
\begin{equation}\label{uno}
3\left(\frac{\dot{a}}{a}\right)^2=8\pi\rho,
\end{equation}
\begin{equation}\label{dos}
3\frac{\ddot{a}}{a}=-4\pi(\rho+3p),
\end{equation}
where $a(t)$ is the scale factor, which can be used to define the Hubble parameter $H=\dot{a}/a$, $\rho$  and $p$ are the energy density and the pressure of the fluid, respectively. Eq. (\ref{dos}) shows that, in order to obtain an accelerating universe, the dark energy must have an energy density and pressure such that $\rho+3p<0$, violating at least the strong energy condition. If one wants to minimise the strange character of the dark energy, then $w>-1$ ($w=p/\rho$) could be imposed, but it must be noted that such restriction is not based in direct observations but in theoretical wishes.

In fact, as we have already mentioned in the introduction, the observational data indicates that $w$ must be around $-1$ and, therefore, values less than $-1$  are not excluded. In that case the fluid is called phantom energy \cite{Caldwell:1999ew} and violates even the dominant energy condition.

The limiting case, $w=-1$, is also possible and is equivalent to the
introduction of a positive cosmological constant. We shall not pay
much attention to this case because it is the well-known de Sitter
solution and more importantly, the cosmological constant would be
accreted neither by black- nor by 
worm-holes, as it could be expected and we show in the next sections.

In this section we present the simplest dark and phantom energy
models, where the equation of state parameter is 
considered to be closely constant. As we shall see, such a
phantom energy model implies the occurrence of 
a big rip \cite{Caldwell:2003vq}. Since it could seem that phantom
energy implies the occurrence of a big rip 
singularity, we want to show that such a singularity is not an inherent 
property of that fluid. So we shall consider Phantom Generalized Chaplygin Gas (PGCG) models, in order to clarify that phantom models could present no 
future singularities \cite{Bouhmadi-Lopez:2004me} or future singularities of other 
kinds \cite{BouhmadiLopez:2006fu,BouhmadiLopez:2007qb}.

\subsection{Quintessence with a constant equation of state parameter.}
The most popular candidate to describe dark energy, allowing a dynamic
evolution for this unknown component, is the quintessence 
model. In this model a spatially homogeneous massless scalar field is considered, which can be interpreted as a perfect fluid with negative pressure, taking the equation of state parameter values on the range $-1<w<-1/3$. Supposing that the equation of state parameter is approximately constant, the conservation law of the fluid in a FLRW spacetime, $\dot{\rho}+3H(p+\rho)=0$, can be integrated to obtain
\begin{equation}\label{tres}
p=w\rho=w\rho_0\left[a(t)/a_0\right]^{-3(1+w)},
\end{equation}
which can be introduced in Eq.~(\ref{uno}) leading 
\begin{equation}\label{cuatro}
a(t)=a_0\left(1+\frac{3}{2}(1+w)C(t-t_0)\right)^{2/[3(w+1)]},
\end{equation}
with $C=\left(8\pi\rho_0/3\right)^{1/2}$ and the subscript $0$ denoting the value at the current time $t_0$. Therefore, a universe described with that model would accelerate forever, decreasing the dark energy density in the process.

\subsection{Phantom quintessence with a constant equation of state parameter.}
Phantom energy can be considered to be a fluid with an equation of
state parameter less than $-1$ which would, therefore, violate not
only the strong energy condition but also the dominant energy
condition\footnote{It must be noted that if we want to express the
phantom fluid by using a scalar field like in the quintessence dark
energy case, then it will possess a negative kinetic term.}. But,
since the observational data suggests that it could be responsible for
the current accelerating expansion therefore, such a pathological fluid should be seriously considered as the possible dominating matter content of our universe.

Analogous to the quintessence case, one can easily obtain an expression for the scale factor in this model considering that $w$ is approximately constant. One has
\begin{equation}\label{cinco}
a(t)=a_0\left(1-\frac{3}{2}(|w|-1)C(t-t_0)\right)^{-2/[3(|w|-1)]},
\end{equation}
with $\rho=\rho_0\left[a(t)/a_0\right]^{3(|w|-1)}$ and $C$ taking the already mentioned value. Therefore, the scale factor~(\ref{cinco}) increases with time even faster than the scale factor of a de Sitter universe (which has an exponential behaviour) up to
\begin{equation}\label{seis}
t_{br}=t_0+\frac{2}{3(|w|-1)C}>t_0.
\end{equation}
At this time both the scale factor and the energy density of the fluid
blow up in which is known as the big rip singularity \cite{Caldwell:2003vq}. If our
Universe is described by this model, then an observer located on
the Earth would see how progressively it could be ripped apart the galaxies, the stars, our solar system and finally, the atoms and nuclei, up until the moment when every component of the universe would be out of the Hubble horizon of the other components.

\subsection{Phantom Generalized Chaplygin Gas.}
It could seem that the consideration of a universe filled with phantom
energy implies the occurrence of a future big rip singularity, but
this is not necessarily the case. We support this claim with the
example 
taken from the Phantom 
Generalized Chaplygin Gas (PGCG) \cite{BouhmadiLopez:2006fu,BouhmadiLopez:2007qb,Bouhmadi-Lopez:2004me}.

The Generalized Chaplygin Gas (GCG) is a fluid with an equation of state of the form \cite{Kamenshchik}
\begin{equation}\label{siete}
p=-\frac{A}{\rho^{\alpha}},
\end{equation}
where A is a positive constant and $\alpha>-1$ is a parameter. In the
particular case $\alpha=1$ we recover the equation of state of a
Chaplygin gas.
Inserting Eq.~(\ref{siete}) in the conservation of the energy momentum tensor, one obtains
\begin{equation}\label{ocho}
\rho=\left(A+\frac{B}{a^{3(1+\alpha)}}\right)^{\frac{1}{1+\alpha}},
\end{equation}
with $B$ a constant parameter. It can be seen that, maintaining $A>0$, such a fluid fulfils the dominant
energy condition for $B>0$ and it is violated otherwise. The rather strange equation of state expressed through Eq.~(\ref{siete}) has been considered firstly in cosmology because the GCG could reproduce a transition from a dust dominated universe at early time to de Sitter behaviour at late time.

On the other hand, PGCG \cite{BouhmadiLopez:2006fu,BouhmadiLopez:2007qb,Bouhmadi-Lopez:2004me,Khalatnikov:2003im} are fluids with an equation of state of the GCG type, Eq.~(\ref{siete}), with the parameters $A$, $B$ and $\alpha$ taking values in intervals such that the energy density and pressure of the fluid fulfil the requirements $\rho>0$ and $p+\rho<0$. It can be seen that such requirements lead to four classes of PGCG with
\begin{itemize}

\item type I: $A>0$, $B<0$ and $1+\alpha>0$.

\item type II: $A>0$, $B<0$ and $1+\alpha<0$.

\item type III: $A<0$, $B>0$ and $(1+\alpha)^{-1}=2n>0$.

\item type IV: $A<0$, $B>0$ and $(1+\alpha)^{-1}=2n<0$.

\end{itemize}
We want to emphasise that when PGCG is considered, the sign of the parameters $A$ and $B$ must not to be necessarily positive and also $\alpha$ can be bigger or less than $-1$.

It can be seen that for type I the scale factor is bounded from below
by $a_{{\rm min}}=|B/A|^{1/[3(1+\alpha)]}$ and, therefore, it takes 
values in the interval $a_{{\rm min}}\leq a<\infty$, which corresponds
to $0\leq \rho<|A|^{1/(1+\alpha)}$, approaching the energy 
density a finite constant value when the scale factor tends to
infinity. The Friedmann Eq.~(\ref{uno}) can be analytically
integrated for the energy density (\ref{ocho}) to lead a functional 
dependence of the cosmic time and the scale factor in terms of
a hypergeometric series \cite{Bouhmadi-Lopez:2004me}. Since this
expression is rather complicated and can be found in the literature,
\cite{Bouhmadi-Lopez:2004me}, we consider that it 
is enough to comment that the resulting expression implies that
when the scale factor diverges the cosmic time also blows up, that
is, there is no a singularity at a finite time in the future. The
future normal behaviour is also indicated by the Hubble parameter
which approaches a constant finite non-vanishing value for large scale
factors and, therefore, we can conclude that the late time evolution of
such a model is asymptotically de Sitter. It can be seen
\cite{BouhmadiLopez:2007qb} that 
the type III model has a similar evolution for late times than the
type I, although both behaviours can differ greatly at early
times\footnote{The quantised $\alpha$ parameter in the type IV model
  eliminates possible past singularities 
that could appear in type I. A discussion about the evolution of these models at early times is out in the scope of the present chapter, so we refer the interested reader to Ref.~\cite{BouhmadiLopez:2007qb}. }.

Although we have just discussed that the consideration of phantom
models could avoid the occurrence of a future big rip singularity,
another kind of doomsday could appear in phantom models. In order to
show this fact, let us consider the type II PGCG. In this case the
scale factor is bound from above by $a_{{\rm
    max}}=|B/A|^{1/[3(1+\alpha)]}$ taking, therefore, values in the
interval $0<a\leq a_{{\rm max}}$ which correspond to
$A^{1/(1+\alpha)}\leq\rho<\infty$. As in the type I an analytical
expression for the scale factor depending on the cosmic time can be
found in terms of hypergeometric series
\cite{BouhmadiLopez:2007qb}. Such an expression can be approximated
close to the maximum scale factor value and  
inverted leading to \cite{BouhmadiLopez:2006fu,BouhmadiLopez:2007qb}
\begin{equation}\label{nueve}
a\simeq a_{\rm{max}}\left\{1-\left(\frac{8\pi}{3}\right)^{\frac{1+\alpha}{1+2\alpha}}\left[\frac{1+2\alpha}{2(1+\alpha)}\right]^{\frac{2(1+\alpha)}{1+2\alpha}}A^{\frac{1}{1+2\alpha}}
|3(1+\alpha)|^{\frac{1}{1+2\alpha}}(t_{\rm{max}}-t)^{\frac{2(1+\alpha)}{1+2\alpha}}\right\},
\end{equation}
which implies that the cosmic time elapsed since the universe has a
given scale factor $a$ (closed to $a_{{\rm max}}$) until it reaches
its maximum value is finite, i.e., $t_{{\rm
    max}}-t<\infty$. Therefore, this model would end at a finite
singularity where its energy density blows up whereas the scale factor
remains finite, called big freeze singularity
\cite{BouhmadiLopez:2006fu}. The type IV model would have a similar
behaviour and it exhibits a singularity of the same kind, \cite{BouhmadiLopez:2007qb}. So, both models describe a universe which would expand accelerating until it
freezes its
evolution at a finite time where it is infinitely full of phantom energy.

It must be pointed out that from a classical point of view, as such we are considering through this chapter, a singularity would break down the spacetime. Nevertheless, it has been argued that the consideration of quantum effects could avoid the big rip \cite{Nojiri:2004pf} and the big freeze singularity \cite{Bouhmadi}.

\section{Dark energy accretion onto black holes.}

As we have pointed out in the previous section, dark energy is filling our Universe and, therefore, it could be expected that it would  interact with different cosmological
objects like black- and worm-holes. This section is dedicated to the
study of the accretion process onto black holes by using the most
accepted model dealing with this phenomenon. Nevertheless, it must be
pointed out that some controversy has originated around this
method and as such other approaches have been proposed, which will be discussed in Sec.~V.

The standard method treating the dark energy accretion onto black holes was firstly presented by Babichev et al. \cite{Babichev:2005py} and is based on the consideration of a black hole described by the Schwarzschild metric, surrounded by a perfect fluid which represents dark energy. In such a framework they considered the zero component of the energy-momentum conservation equation and the projection of this equation along the four-velocity, to derive the dynamical evolution of the black hole mass. We want to summarise a generalisation of this method, presented in Ref.~\cite{MartinMoruno:2006mi}, which allows
an internal nonzero energy-flow component $\Theta_{0}^{r}$ by the consideration of the simplest
non-static generalisation of the Schwarzschild metric, in which the
black hole mass can depend generically on time. That is, the
metric can be given by

\begin{equation}
{\rm d}s^2=\left(1-\frac{2M(t)}{r}\right){\rm d}t^2-\left(1-\frac{2M(t)}{r}\right)^{-1}{\rm d}r^2- r^2\left({\rm d}\theta^2+\sin^2{\rm d}\phi^2\right),
\end{equation}
where $M(t)$ is the black hole mass.

The zero component of the conservation law for energy-momentum tensor and its projection
along the four-velocity can be integrated in the radial coordinate
considering a surrounding perfect fluid. On the other hand, it is
known that the rate of change of the black hole mass
due to accretion of dark energy can be derived by
integrating over the surface area the density of momentum $T_0^r$. Taking into account these considerations one
gets the following equation \cite{MartinMoruno:2006mi}

\begin{equation}
\dot{M}=4\pi A_{M} M^2\left(p+\rho\right)e^{-\int_\infty^r f\left(r,t\right){\rm d}r},
\end{equation}
relating the temporal rate of change of the mass with the pressure and 
energy density of the perfect fluid which is considered to describe dark energy.

For the relevant physical case of an asymptotic observer, i.e. $r\rightarrow
\infty$, the previous equation simplifies to

\begin{equation}\label{eq:negrofinal}
\dot{M}=4\pi A_{M} M^2\left(p+\rho\right),
\end{equation}
where $A_{M}$ is a positive constant of order unity. It must be pointed out that this expression is the same as that obtained by Babichev et al. \cite{Babichev:2005py} using the usual Schwarzschild metric, with the difference that in the current case it is only valid for asymptotic observers.

It can be noted that Eq.~(\ref{eq:negrofinal}) shows that the black hole mass, and with it its size, must decrease when the
black hole accretes a fluid which violates the dominant energy condition,  i.e. a phantom fluid. This would
increase when the dominant energy condition is preserved and it remains constant in the cosmological constant case, since a cosmological constant can be modelled by a perfect fluid with $p+\rho=0$.
It must be emphasised that Eq.~(\ref{eq:negrofinal}) has been obtained for a general
perfect fluid, therefore it would be valid in a great variety of dark energy models,
since a load of them are based on a perfect fluid.

Since accretion of dark energy onto black holes would 
increase the black hole size in dark 
energy models fulfilling the dominant energy condition, an interesting
question is whether this growth could be large enough 
to that the black hole might engulf the whole universe. In order to find a possible answer for this question, one can take into account the Friedmann equations to integrate Eq.~(\ref{eq:negrofinal}), obtaining the temporal evolution of the black hole mass. That is, \cite{Yo}
\begin{equation}\label{M4D}
M=\frac{M_0}{1+\sqrt{\frac{8\pi}{3}}A_{M}M_0\left(\rho^{1/2}-\rho_0^{1/2}\right)}.
\end{equation}
This equation can also be expressed in terms of the Hubble parameter in the following way
\begin{equation}\label{M4DH}
M(t)=\frac{M_0}{1+D M_0\left[H(t)-H_0\right]}.
\end{equation}
Therefore, as mentioned in Ref.~\cite{MartinMoruno:2006mi}, a black
hole capable of engulfing the whole universe as shown in a model with an
equation of state parameter bigger than minus one, should have a
current mass which, roughly speaking, is bigger than all the matter
content of the current observable universe, making 
the occurrence of such a phenomenon impossible.

In the following subsections we consider the models reviewed in the previous section to study the evolution of a black hole
living in a universe filled with dark energy.

\subsection{Application to a quintessence  model.}

We assume that dark energy is modelled by a quintessence model satisfying the equation
of state (\ref{tres}) with $w>-1$ constant. As we have already mentioned, Eq.~(\ref{eq:negrofinal}) implies
that the rate of mass change is positive in this model, therefore the black hole mass grows along time due
to accretion of quintessence. In order to obtain the dynamical behaviour of the black hole mass, one can introduce the equation of state (\ref{tres}) in Eq.~(\ref{M4D}) which leads to

\begin{equation}\label{eq:mnegroquintaesencia}
M=\frac{M_0 \left[1+\frac{3}{2}\left(1+w\right)C\left(t-t_0\right)\right]}{1+\frac{3}{2}\left(1+w\right)C\left(t-t_0\right)-4\pi A_{M}\rho_0 M_0\left(1+w\right)\left(t-t_0\right)} .
\end{equation}
There are something very interesting in this expression for the black
hole mass, because it allows the occurrence of a 
bizarre fate for our universe as we have already pointed out. That is,
Eq.~(\ref{eq:mnegroquintaesencia}) expresses 
the possibility that our universe might be engulfed by a black hole, since
accretion of dark energy could make the mass of the black hole
increase so 
quickly
as to yield a black hole size that would
eventually exceed the size of the universe in a finite cosmic time. In fact, the time 
in which the black hole might reach a infinite size would be
\begin{equation}
t_{bs}=t_0+\frac{1}{\left(1+w\right)\left(4\pi A_{M}\rho_0 M_0-\sqrt{6\pi\rho_0}\right)}.
\end{equation}
This time is finite but, although a universe filled with
quintessence has no future singularity,
present observational data seems to imply that $w$ is not constant and
suggests values less than -1, making it unlikely that the occurrence of
the considered bizarre phenomenon  at any time in the far future would
happen, at least in principle. However, if $\dot{w} > 0$ then the black holes would undergo a larger growth due to accretion of dark energy.

Nevertheless, if one studies in deeper detail
Eq.~(\ref{eq:mnegroquintaesencia}), then one obtains
(see Ref.~\cite{MartinMoruno:2006mi}) that in order to have a black
hole able to reach an infinite mass in an infinite 
time, this black hole must possess an initial mass such as 
$(8\pi\rho_0/3)^{1/2}A_{M}M_0=1$ which means, taking 
into account the observational data, $M\sim
10^{23}M_\odot$ (where $M_\odot$ is the Sun's
mass). Therefore, in order to have an infinitely large black 
hole in a finite time in the future, the current black hole mass
should be bigger than $10^{23}M_\odot$, which is an extremely
large value even for black holes in the galaxies centres. Even more,
one can estimate the current matter content of
the universe assuming that the observable Universe expands at the
speed of light, obtaining a total
of $10^{23}$ starts \cite{MartinMoruno:2006mi}. Therefore, it seems
that in order to have a black hole able to engulf
the whole universe, it should have a current mass equal to the mass of
all the observable universe, which would not
be possible.

Finally, we want to point out that, even in the case that the
accretion phenomenon of dark energy onto black holes 
could not produce cosmological consequences in terms of a catastrophic
end, it could help in the determination of
the correct dark energy model. So, if astronomers were able, in practice, to observe a growth bigger than expected of those black holes
living in the centre of the galaxies, then this could
be an observational measure of the effects of dark energy with $w>-1$.

\subsection{Application to a phantom quintessence model.}

As we have already pointed out, the observational data not only allows
that the equation of state parameter 
takes a value less than $-1$ but even they seem to suggest it,
acquiring therefore, special interest in the study of 
the evolution of a black hole in a universe filled with phantom energy.
So, now we consider $w<-1$ and constant, consequently
(\ref{eq:negrofinal}) shows that the black hole mass decreases with time.
More precisely, inserting Eqs.~(\ref{tres}) and (\ref{cinco}) in Eq. (\ref{M4D}), we
get an accurate expression of the evolution of the black hole mass, i.e.

\begin{equation}\label{eq:mnegrophantom}
M=\frac{M_0 \left[1-\frac{3}{2}\left(|w|-1\right)C\left(t-t_0\right)\right]}{1-\frac{3}{2}\left(|w|-1\right)C\left(t-t_0\right)+4\pi A_{M}\rho_0 M_0\left(|w|-1\right)\left(t-t_0\right)} .
\end{equation} 
Taking into account that in a universe filled with phantom energy with $w$
constant a big rip singularity will take place in the future, one can introduce
the time of occurrence of the big rip, $t_{br}$, in
(\ref{eq:mnegrophantom}) to get that the black hole mass vanishes at
$t_{br}$,
independently of the current black hole mass, $M_0$; that is, all
black holes disappear at the big rip \cite{Babichev:2005py}. It can be noted, by
inspection of Eq.~(\ref{M4DH}), that this is not only an interesting
property of this phantom model but it can be found 
in all models which present a singularity with a divergence of the Hubble parameter at a finite time in the future.

Finally, the decrease of black holes due to the phantom energy
accretion phenomenon could provide us with a possible 
observational test of these models. Therefore, if future observations
of black holes in the centre of 
galaxies (or other possible black holes) indicate a growth of
those objects less than expected, then it could 
be associated to accretion of phantom energy,  providing us with another measure able to discriminate between different dark energy models, which would complete
those that come from GRB
\cite{Wang:2009zzq,Ghirlanda:2006ax}, supernova, or other
observational data.

\subsection{Application to a Generalized Chaplygin model.}

Let us now study the evolution of a black hole in a universe
filled with a Generalized Chaplygin Gas. Since the consideration of dark energy modelled by some kinds of Phantom
Generalized
Chaplygin Gas could prevent the occurrence of a big rip \cite{Bouhmadi-Lopez:2004me}, one
could expect the avoidance of the weird behaviours that appear in
quintessence or  phantom models in these frameworks.  In order to see if that is the case, one must take into account Eq. (\ref{M4D}).

As we have mentioned in Sec.~II, a Generalized Chaplygin Gas, with $A>0$ and $\alpha>-1$, preserves dominant energy
condition when the parameter\footnote{$B$ is the constant parameter that
  appears in the energy momentum tensor conservation law
  (\ref{ocho}) for a GCG.} $B>0$ and violated otherwise. So when $B>0$ the black
hole mass increases with cosmic time up to a constant value, but there
are a set of parameters where if $\alpha$ is close to $-1$ then it would seem that the
black hole mass could eventually exceed the size of the universe at
finite time in the future \cite{Jimenez Madrid:2005gd}. 
When dominant energy condition is violated, $B<0$, black hole
mass decreases along cosmic time, tending to a nonzero constant value,
therefore black holes do not disappear in this model.

On the other hand, it can be seen \cite{Yo} that performing a deeper
study of the mentioned four types of 
PGCG (where the sign of $A$ and the range on $\alpha$ is not previously fixed) at late times, where the phantom fluid would drive the dynamical evolution of the universe, the results can be summarised as follows:

\begin{itemize}
\item Type I and III. Black holes decrease with time, where
the mass tend to a nonzero value when the time goes to infinity.
\item Type II and IV. Black holes masses decrease, but now all
black holes disappear at big freeze, i.e., the mass of all black holes
tend to zero when the universe reaches the big freeze singularity with
independence of their initial mass (as it should be expected by inspection of equation (\ref{M4DH})).
\end{itemize}

 To end this section, we consider that  it would be quite interesting to explore the region of parameter space 
$(\alpha, A , H_0 , \Omega_K ,\Omega_\phi)$ allowed by current
  observations in order to determine 
whether there exists any allowed sections leading to a big freeze or a
big rip. However, all available analyses
\cite{Bertolami:2004ic,Biesiada:2004td,Zhu:2004aq,Colistete:2005yx,Lu:2009zzf,delCampo:2009cz,Li:2009br,Wang:2009zzq,Wu:2007zz,Lu:2008zzb} 
are restricted to the physical region where
no dominant energy condition is violated.
Therefore, the section described by the interval implied by a PGCG necessarily is outside the analysed regions. One has
to extend the investigated domains to include values of parameter
$A>1$, $A<0$ or $\alpha<-1$ to probe the parameters space where
the dominant energy condition is violated.

\subsection{Consideration to other black holes.}

Up to now, we have shown the evolution of a Schwarzschild black hole
in a universe filled with dark energy. In this subsection, we study the
accretion of dark energy onto charged or rotating black holes, to show whether
charge or angular momentum have some influence in their evolution.

 Let us continue by considering another black hole metric in order to understand
better the application of the dark energy accretion mechanism. 
In \cite{Babichev:2008jb}, Babichev et al. apply a generalisation of
the accretion formalism to a Reissner-Nordsr\"om black hole. In this
case, the metric is given by

\begin{equation}
{\rm d}s^2=\left(1-\frac{2M}{r}+\frac{e^2}{r^2}\right){\rm d}t^2-\left(1-\frac{2M}{r}+\frac{e^2}{r^2}\right)^{-1}{\rm d}r^2
- r^2\left({\rm d}\theta^2+\sin^2\theta{\rm d}\phi^2\right),
\end{equation}
where $m^2>e^2$. It can be noted that if $m<|e|$, then the solution would represent a naked
singularity, corresponding $m=|e|$ to the extreme case.
By integrating the conservations laws for momentum-energy and its projection
along four-velocity for the case for a perfect fluid, and taking into
account that the rate of change of the black hole mass
due to accretion of dark energy can be derived by
integration over the surface area the density of momentum $T_0^r$,
Babichev et al. get again the same expression (\ref{eq:negrofinal})
which relates the temporal rate of change of the black hole mass to the 
pressure and the energy density of the perfect fluid.
So, also in a Reissner-Nordsr\"om black hole, the black hole mass
increases when it is accreting dark energy holding the dominant energy
condition
and its mass decreases when phantom energy is considered in the
accretion process.

At this point the next question arises again, if phantom energy is getting
involved then black hole mass decreases, vanishing at big rip
singularity. Therefore, since the electric charge $e$  remains constant due to phantom
energy accretion, then in a finite time, the black hole must reach the
extreme case, transforming the black hole into a naked singularity.
Nevertheless, the authors of Ref. \cite{Babichev:2008jb} perform a
more detailed study 
about this possible transformation in a naked singularity, concluding that there is no accretion of the perfect fluid
onto the Reissner-Nordstr\"om naked singularity when $m^2<e^2$ and that, in
this situation, a static atmosphere of the fluid around the
naked
singularity  would be formed.

It must be emphasised that, although when one is considering cases far
from the extreme case, the back reaction can be 
neglected and the perfect fluid approximation appears to be valid, it
seems that this approximation breaks down 
close to the extremal case, where one has to take into
account the back reaction of the perfect fluid onto the background metric. Even more, the same consideration about the avoidance of transformation of a black hole into a naked singularity, can also be applied to a Kerr black
hole. Nevertheless, if the back reaction does not prevent the process of phantom accretion onto a
charged black hole or rotating black hole, then this process could be
a way to violate the cosmic censorship conjecture \cite{Penrose}.

\section{Dark energy accretion onto wormholes.}

The first solution of the Einstein's equations describing a traversable wormhole was found by Morris and Thorne in their seminal work \cite{Morris:1988cz}. That solution, obtained under the assumption of staticity and spherical symmetry, describes a throat connecting two asymptotically flat regions of the spacetime without any horizon and can be expressed as
\begin{equation}\label{diez}
 {\rm d}s^2=-e^{2\Phi(r)}{\rm d}t^2+\frac{{\rm d}r^2}{1-K(r)/r}+
 r^2\left({\rm d}\theta^2+\sin^2{\rm d}\varphi^2\right),
\end{equation}
where $\Phi(r)$ and $K(r)$ are the shift and
shape functions, respectively, both tending to a constant value when the radial coordinate $r\rightarrow\infty$ in order to have asymptotic flatness. It must be noted that, in these coordinates, two coordinate patches are needed to cover the two
asymptotically flat regions, each with $r_0\leq r\leq\infty$, with $r_0$ the minimum radius which corresponds to the throat radius, where $K(r_0)=r_0$.

It can be seen \cite{Morris:1988cz} that solution (\ref{diez}) must
fulfil some additional requirements in order to describe a traversable
wormhole. In particular the outward flaring condition imposes
$K'(r_0)<1$ what, through the Einstein's equations, implies that
$p_r(r_0)+\rho(r_0)<0$ (where here $p_r$ denotes the radial component
of the pressure). Therefore, the wormhole must be surrounded by some
material with unusual characteristics, called exotic matter, which
could 
lead to the neglect of such spacetime.

Nevertheless, as we have already mentioned in the introduction, the
discovery of the current accelerated expansion of 
the Universe and the consideration of phantom energy as a possible
candidate for its origin has produced a more natural 
consideration of the properties of exotic matter, since it seems that
phantom energy could be precisely the exotic stuff 
which supports wormholes. Gonzalez-Diaz \cite{GonzalezDiaz:2004vv}
considered that similarly to black holes accrete dark energy,
wormholes could accrete phantom energy producing a great increase of
their size, in such a way that the 
size of a wormhole could be infinitely larger before the universe reaches the big rip singularity. Such a process would produce the moment that the size of the wormhole equals the size of the universe, the universe boards itself in a travel through the wormhole, called big trip.
Even more, the notion of phantom energy has been extended to
inhomogeneous spherically symmetric spacetimes showing that it can be
in fact the exotic material which supports wormholes
\cite{Sushkov:2005kj}, which backs up the mentioned idea.

In order to study such a process one can follow a similar method to the one used by Babichev et al. for the black hole case. So, considering the non-static generalisation of Eq.~(\ref{diez}) obtained by the consideration of an arbitrary dependence of the shape function on the time, $K(r,t)$, and an energy momentum-tensor of a perfect fluid, one can find the equivalent of Eq.~(\ref{eq:negrofinal}) for the wormhole case, that is the temporal mass rate evolution as measured by an asymptotic observer, which is \cite{GonzalezDiaz:2007gt}
\begin{equation}\label{once}
\dot{m}=-4\pi Qm^2(p+\rho),
\end{equation}
with $Q$ a positive constant. That expression shows that the wormhole
mass, and with it, its size must increase when the wormhole accretes
phantom energy, it decreases by the accretion of dark energy,
remaining constant in the cosmological constant case. It must be
remarked that in the achieving of Eq.~(\ref{once}) no assumption about
the possible dependence on the energy density or on the
pressure of the fluid have been done, allowing an arbitrary 
dependence with the time and with the radial component, therefore,
such an equation is general and take into account 
the possible back reaction in an asymptotically flat wormhole
spacetime\footnote{It must be emphasised that, whereas in the 
case of the study of the dark energy accretion onto black hole
phenomenon, the solution is not able to take into account 
the back reaction of the spacetime, in this case we are treating with
a non-vacuum solution and allowing arbitrary 
time dependence on the involved functions and, therefore, up to now we are taking into account any possible back reaction.}.

If one now considers as an approximation that the fluid which surrounds the wormhole is a cosmological one, i.e., an homogeneous and isotropic fluid fulfilling the Friedmann equations (\ref{uno}) and (\ref{dos}) and the conservation law, then Eq.~(\ref{once}) can be integrated to obtain \cite{Yo,Jimenez Madrid:2005gd}
\begin{equation}\label{doce}
m(t)=\frac{m_0}{1-Qm_0\left[H(t)-H_0\right]}.
\end{equation}

This expression shows that in phantom models, where $H(t)$ is an
increasing function, the wormhole throat could become 
infinitely big if the Hubble parameter reaches the value
$H_*=H_0+1/(Qm_0)<\infty$ at some time $t_*$ in the future. 
It can be seen that this would be the case at least in models which
show a future singularity in a finite time in 
the future characterised by a divergence of the Hubble parameter,
because in those models 
one has $H_0<H_*<H_{{\rm sing}}=\infty$ and, since the Hubble parameter is
a strictly increasing and continuous function 
before the time of the singularity, this implies $t_0<t_*<t_{{\rm sing}}$. Therefore, the size of a wormhole would be bigger than the size of the universe before the occurrence of the future singularity in all models possessing a future singularity where the Hubble parameter blows up, i.e., in such models the universe would travel through a big trip.

In this section we show the implications of the phenomenon of dark and
phantom energy accretion onto wormholes in the models presented in
Sec.~II. That procedure lead, as it is expected, to the decrease of
the wormhole size when it accretes dark 
energy with $w>-1$ and to a growth of the wormhole mouth in phantom
cases. Even more, the big rip and big freeze singularities, in the
corresponding models, can be avoided by a big trip phenomenon
since, although these singularities present 
a different behaviour of the scale factor, at both singularities the Hubble parameter blows up.

\subsection{Application to a quintessence model.}

Let us consider that the dark energy is modelled by a quintessence model, satisfying the equation
of state (\ref{tres}) with $w>-1$ constant. Eq.~(\ref{once}) tells us
that the rate of mass change is negative, so the wormhole mass would
decrease  with time due
to the accretion of quintessence. Furthermore, taking into account the equation of state (\ref{tres}),
one can solve (\ref{once}) getting the following expression which relates the wormhole mass
to the cosmic time \cite{GonzalezDiaz:2005yj},

\begin{equation}\label{eq:mgusanoquintaesencia}
m=\frac{m_0}{1+\frac{4\pi Q\rho_0 m_0\left(1+w\right)\left(t-t_0\right)}{1+\frac{3}{2}\left(1+w\right)C\left(t-t_0\right)}} .
\end{equation} 
This expression shows us how a wormhole loses mass due to the accretion of
quintessence.
Even more, if due to any additional hypothetical process this wormhole would have a macroscopic size, then it would be subjected to chronology protection
\cite{Hawking:1991nk}; therefore, vacuum polarisation created particles would
catastrophically accumulated on the chronology horizon of the wormhole,
letting the corresponding normalised stress-energy tensor to diverge which, at the end of the day, would imply the disappearance of the wormhole.

\subsection{Application to a phantom quintessence model.}

Now, we are interested in study the evolution of a
wormhole in a universe filled with phantom energy with
$w<-1$ constant. In order to obtain the temporal evolution of the wormhole, one can introduce the equation of state of
phantom energy into  the r.h.s. of Eq. (\ref{once}), getting
\cite{GonzalezDiaz:2005yj},

\begin{equation}\label{eq:mgusanophantom}
m=\frac{m_0}{1+\frac{4\pi Q\rho_0 m_0\left(|w|-1\right)\left(t-t_0\right)}{1+\frac{3}{2}\left(|w|-1\right)C\left(t-t_0\right)}}.
\end{equation} 
This expression implies 
that the exotic mass of the wormhole diverges at the time

\begin{equation}
t_{bt}=t_0+\frac{t_{br}-t_0}{1+\frac{8\pi Qm_0a_0^{3\left(|w|-1\right)/2}}{3C}},
\end{equation}
where $t_{br}$ is the finite time at which the big rip singularity takes place.
Since the wormhole size diverges before that the universe reaches the big rip singularity, it would be a previous time at which the size of the wormhole would be bigger than the universe, being at this time where properly starts the travel of the universe through the wormhole.

The huge growth of the wormhole throat poses the following two
problems. On the one hand, since the wormhole
spacetimes are usually considered to be asymptotically flat, when the
wormhole increases more than the universe
it is impossible to place the wormhole on this universe. On the other hand, since the universe is travelling through the wormhole, one can ask where is the universe travelling to? The solution of these equivalent problems requires the consideration of a multiverse scenario. In such a framework  
the wormhole could be re-infixed in another universe where the
wormhole would be asymptotically flat to, giving also a final
destination to the universal travel\footnote{It is worth noticing 
that in models showing one big trip, as it is the considered
case, the universe would travel through the time of the 
arrival universe being, in this case, not a proper time travel. On the
other hand, in models which present more than one 
big trip phenomenon the consideration of a multiverse framework would
be not more necessary, since the wormhole mouth at 
the moment that it is bigger than the universe could be connected to
the other infinitely large wormhole mouth, travelling 
in this case the universe along its own time from future to past. The
reader interested on this topic is advised to 
consult Ref. \cite{Yurov:2006we}.}.

\subsection{Application to a Generalized Chaplygin Gas model.}

Finally, we will study the evolution of a wormhole in the case of
a universe filled with a Generalized Chaplygin Gas.
Since type I and III of PGCG avoid the occurrence of a future singularity \cite{BouhmadiLopez:2007qb,Bouhmadi-Lopez:2004me}, it is of special interest to study the possible occurrence of a big trip phenomenon in these models. Following this line of thinking, in Ref.~\cite{Jimenez Madrid:2005gd} it is analysed the phantom energy accretion phenomenon onto wormholes when the phantom energy is modelled by a type I PGCG. In order to perform this study, let us temporarily fix $A>0$ and $\alpha>-1$, therefore, solving Eq.~(\ref{doce}) for the equation of state of a GCG, one obtains

\begin{equation}\label{eq:mgusanochaplygin}
m=\frac{m_0}{1-Qm_0\sqrt{\frac{8\pi}{3}}\left(\rho^\frac{1}{2}-\rho_0^{1/2}\right)}.
\end{equation}

For the case where the dominant energy condition is violated, i.e. $B < 0$, we obtain that $m$ increases with time and
tends to a maximum, nonzero constant value. If
the dominant energy condition is assumed to be hold, i.e. $B > 0$, 
then $m$ decreases with time, with $m$ tending to  nonzero constant
values.

It can be seen that, when the cosmic time goes to infinity, then the exotic mass of wormhole
approaches to

\begin{equation}
m=\frac{m_0}{1-Qm_0\sqrt{\frac{8\pi}{3}}\left(A^\frac{1}{2\left(1+\alpha\right)}-\rho_0^{1/2}\right)},
\end{equation}
that is a generally finite value both for $B > 0$ and $B < 0$. Thus, it could be thought that
the presence of a Generalized Chaplygin Gas prevents the eventual
occurrence of the big trip phenomenon. However, such a conclusion cannot be guaranteed as the size of the wormhole
throat could still exceed the size of the universe during its previous
evolution. The
question is whether the wormhole would grow rapidly enough or not to engulf the universe during the
evolution to its final classically stationary state. To avoid a big
trip one needs that the radius of the wormhole does not exceed the size
of the universe. It can be checked \cite{Jimenez Madrid:2005gd} that
GCG generally prevents the occurrence of a big trip when  $\alpha$ does
not reach values sufficiently close to $-1$, but when $\alpha$ is
inside the interval

\begin{equation}
-1<\alpha<\frac{\ln A}{\ln\left(\sqrt{\frac{3}{8\pi}}\frac{1}{m_0 D}+\rho_0^{1/2}\right)^2}-1,
\end{equation}
a big trip would still take place.

It is worth noticing that, when there is no big trip phenomenon, the wormhole size tends to become
constant at the final stages of its 
evolution being a rather a macroscopic object. So, the wormhole at this stage would be subjected
to chronology protection \cite{Hawking:1991nk} and
vacuum polarisation created particles would catastrophically accumulate on the chronology horizon of the wormhole making
the corresponding renormalised stress-energy tensor to diverge and hence the wormhole would disappear.

On the other hand, one can study the wormholes' evolution living
in a universe with phantom  energy modelled by a type II or IV PGCG
\cite{Yo}, where a future big freeze singularity is predicted. Since
a big 
freeze singularity implies the divergence of the Hubble parameter at
this singularity, as we have mentioned in the introduction of the 
present section, this implies that the wormhole size would blow up before the occurrence of the singularity, implying a big trip phenomenon. That can be easily proved taking into account
Eqs.~(\ref{uno}), (\ref{ocho}) and (\ref{doce}) for type II PGCG, which yields
\begin{equation}\label{masachap}
m(x)=\frac{m_0}{1+\sqrt{\frac{8\pi}{3}}\frac{Q m_0}{A^{\frac{1}{2|1+\alpha|}}}\left[\frac{1}{(1-x_0^{3|1+\alpha|})^{\frac{1}{2|1+\alpha|}}}-\frac{1}{(1-x^{3|1+\alpha|})^{\frac{1}{2|1+\alpha|}}}\right]},
\end{equation}
where $x=a/a_{{\rm max}}$ ($0\leq x\leq1$) and a similar expression can be obtained for type IV replacing $A$ with
$|A|$. In order to study the behaviour of the wormhole mass, one can define the function $F(x)=m_0/m(x)$ which is continuous in the interval $[x_0,1)$. This function takes a value $F(x_0)=1>0$ and tends to minus infinity when $x$ goes to 1 (which corresponds to $a\rightarrow a_{{\rm max}}$), which implies that $F(x)$ vanishes at some $x_*$ with $x_0<x_*<1$. Therefore, $m(x)$ blows up at $x_*$ being the throat size
infinitely large before the universe reaches the big freeze singularity (at $x=1$). So the whole universe will travel through the wormhole before the occurrence of the doomsday.

The results can be summarised as follows
\begin{itemize}
\item Types I and III. The evolution of the wormhole is the same 
  for  GCG.
\item Types II and IV. A big trip phenomenon would prevent the
  expected cosmological doomsday, i.e., the big freeze.
\end{itemize}

\section{Debate and new lines of research.}

In the present chapter we have used methods based in the Babichev et al. one, in order to consider the dark and phantom energy accretion onto black- and worm-holes. Our intention is not to claim that the study of these processes is a closed issue, on the contrary, it remains open up to now and a lot of discussion has been originated in this way. 

In order to point out the shortcomings of the current available
methods which deal with the mentioned accretion phenomenon and 
suggest possible new lines of research. In this section 
we also include in chronological order some comments which have
appeared in the literature supporting, improving or 
criticising the Babichev et al. method
\cite{Babichev:2004yx}. We alternate works regarding the accretion
onto 
black holes with others dealing with the accretion onto wormholes,
because both phenomena can be studied following the same 
procedure. Nevertheless, as it has been and will be pointed out, the
wormhole case is free of some shortcomings which 
affect the black hole one, since the first one can never be
  considered as a vacuum solution (at least if one 
restricts oneself to the traversable wormhole case). 

The first work about accretion of dark energy onto black holes was due
to E.~Babichev, V.~Dokuchaev 
and Yu.~Eroshenko \cite{Babichev:2004yx}. They considered the
spherically symmetric accretion of dark energy onto 
black holes adjusting the analytic relativistic accretion solution
onto the Schwarzschild black hole developed 
by Michel \cite{Michel}, eliminating from the equations the particle
number density. So they obtain the expression 
for the black hole temporal mass rate
\begin{equation}\label{unob}
\dot{M}=4\pi A_{M} M^2\left[\rho_\infty+p\left(\rho_\infty\right)\right],
\end{equation}
showing that the black hole mass could decrease by the
accretion phenomenon. The authors pointed out that such a 
decrease of the black hole size is due to the violation of the
dominant energy condition, since this condition is assumed 
to be fulfilled in the derivation of the black hole non-decrease area
theorem. On the other hand, by 
integrating (\ref{unob}) in a phantom Friedmann universe, they found
that the masses of all black holes tend towards zero 
when the universe approaches the big rip, independently of their initial masses.

Soon after, P.~F.~Gonzalez-Diaz \cite{GonzalezDiaz:2004vv} considered
the spherically symmetric accretion of dark 
and phantom energy onto Morris-Thorne wormholes. He assumed that,
since the mass  of the spherical thin shell of the 
exotic matter in a Morris-Thorne wormhole, $\mu=-\pi b_0/2$ (where $b_0$ is the radius of the spherical wormhole throat),
is approximately just the negative of the amount of the mass required
to produce a Schwarzschild wormhole, then the rate 
of change of the wormhole throat radius should be similar to that
obtained by Babichev 
et al. \cite{Babichev:2004yx} for the black hole mass but with a minus sign, i.e.
\begin{equation}\label{tresb}
\dot{b_0}=-2\pi^2 Q b_0^2\left(1+w\right)\rho,
\end{equation}
with $Q\simeq A_{M}$. He concluded, therefore, that the wormhole throat
should increase by the accretion of phantom 
energy. Even more, he showed, by integrating Eq.~(\ref{tresb}) in a
phantom model with constant equation of 
state parameter, that the wormhole increases even faster than the
universe itself, engulfing the whole universe 
before it reaches the big rip singularity. Therefore, the universe would embarks itself in a big trip.

Later on, P.~F.~Gonzalez-Diaz and C.~L.~Siguenza,
\cite{GonzalezDiaz:2004eu}, obtained that the phantom energy 
accretion onto black holes leads to the disappearance of the black
holes at the big rip even 
when Eq.~(\ref{unob}) is integrated in more complicated models
and Babichev et al. recovered their previous result 
in a more detailed work \cite{Babichev:2005py} where they also included
a deeper study of two dark energy models 
admitting analytical solutions. On the other hand, works containing
relevant implications of the result of Babichev 
et al. were also published during that time, like the influence of the
accretion phenomenon on the black hole and 
phantom thermodynamics, Ref. \cite{GonzalezDiaz:2004eu}, and
the possible survival of black holes at the big rip due to the same
phenomenon which could smooth the big rip 
singularity when quantum effects are taken into account, Ref. \cite{Nojiri:2004pf}.

But the previous mentioned works did not had the final say about this
topic. In 2005 V.~ Faraoni 
and W.~ Israel \cite{Faraoni:2005it} considered the time evolution of
a wormhole in a phantom Friedmann universe 
finding no big trip phenomenon. In that work they commented that the
way in which Gonzalez-Diaz applied the Babichev 
et al. method to the wormhole case in Ref. \cite{GonzalezDiaz:2004vv}
could be wrong, since  $\mu=-\pi b_0/2$ must not to 
be necessarily valid for a time-dependent wormhole embedded in a FLRW
universe and, therefore, the wormhole mass 
time rate due to the accretion phenomenon would not be simply the
analogous negative of the black hole mass time rate.

Soon after, Gonzalez-Diaz \cite{GonzalezDiaz:2005yj} followed a
similar procedure as it has been done by 
Babichev et al. \cite{Babichev:2004yx}, adjusting the
Michel theory to the case of Morris-Thorne 
wormholes, in order to study the dark energy accretion onto
wormholes. He obtained Eq.~(\ref{once}) for the case of 
an asymptotic observer
(which is equivalent to Eq.~(\ref{tresb}) with a re-definition of the
constants). He also claimed that the results obtained 
by Faraoni and Israel \cite{Faraoni:2005it} just take into account the
inflationary effects of the accelerated expansion 
of the universe on the wormhole size (also considered by himself years
ago in another work \cite{GonzalezDiaz:2003pb}) and 
do not include the superposed effects due to the accretion phenomenon, which existence is clarified in that work \cite{GonzalezDiaz:2005yj}.

On the other hand, a number of difficulties related to the big trip
process were treated also by Gonzalez-Diaz 
in Ref. \cite{GonzalezDiaz:2006na}. First, he showed how the
corrections appearing in the expressions of the study of a 
wormhole metric with a non-static shape function applying the Babichev
et al. method \cite{Babichev:2004yx} should 
disappear on the asymptotic limit, coinciding with those corresponding
calculated expressions in the static case in that regime. Even more,
he considered explicitly a metric able to describe 
a wormhole in a Friedmann universe, arguing that it would ultimately imply the occurrence of a big trip phenomenon.
Second, since wormhole spacetimes are usually considered to be
asymptotically flat, then when the wormhole increases more 
than the universe, this object can neither be placed on it nor be
asymptotically flat to it. He proposed 
that in such a situation a multiverse context must be considered, what
would allow to re-insert the wormhole in another 
universe, recovering the meaning of the asymptotic regime where the
accretion process has been calculated. In the third 
place, he considered the possible instability of wormholes due to the
quantum creation of vacuum particles on the 
chronology horizon when the wormhole throat grows at a rate smaller
than or nearly the same as the speed of light. However, 
in a phantom model the accreting wormhole would clearly grow at a rate
which exceeds the speed of light asymptotically 
and so the vacuum particles would never reach the chronology horizon
where they have being created, keeping the wormhole 
stability. Moreover, although it was known that quantum effects could
affect the big rip singularity \cite{Nojiri:2004pf}, 
Gonzalez-Diaz showed that those effects have no influence in the big
trip, which would take place before that singularity. 
In the fourth place, he argued that the big trip phenomenon would not
imply any contradiction with the holographic 
bound, since wormholes able to connect regions after and before the
  big rip extend the evolution of the universe up 
to infinite time.

Following that line of thinking, the authors of
\cite{MartinMoruno:2006mi} applied the Babichev et al. method to the 
simplest non-static generalisation of the Schwarzschild metric, in
order to study the dark energy accretion onto 
black holes with arbitrary accretion rates. Although they are
still using a test fluid approach and, therefore, the 
validity of their result on arbitrary accretion rates is debatable,
the non-static metric is enough to take into 
account internal non-zero energy flow $\Theta_0^r$. As it was suggested by Gonzalez-Diaz in the case of wormholes \cite{GonzalezDiaz:2006na}, a study of the accretion phenomenon using a non-static metric recovers the result obtained in the static case, Eq.~(\ref{unob}), for asymptotic observers.

Later on, Faraoni published ``No ``big trips'' for the universe'',
\cite{Faraoni:2007kx}, where a sceptical attitude about the
big trip phenomenon is adopted, based on some shortcomings of the
works of 
Gonzalez-Diaz \cite{GonzalezDiaz:2004vv,GonzalezDiaz:2005yj,GonzalezDiaz:2006na}
in particular and the method of 
Babichev et al.~\cite{Babichev:2004yx} in general. His principal
objection regarding the mentioned works of Gonzalez-Diaz 
was that the use of a static metric can never produce a non-zero
radial energy flow onto the hole, i.e. static metrics 
always imply $\Theta_0^r=0$. Even more, the solution of Gonzalez-Diaz
(and the corresponding of Babichev et al. in the 
black hole case) cannot be adjusted to satisfy the Einstein's
  equations, as the used conservation laws would 
only strictly correspond to vacuum solutions. On the other hand, he
also showed that if the phantom fluid is modelled 
by a perfect fluid, as it is done in the Babichev et al. method and
its application to wormholes, the proper radial 
velocity of the fluid is $v\sim a^{3(1+w)/2}$ which vanishes at the big rip stopping the accretion phenomenon.

Soon after, Gonzalez-Diaz et al.,
\cite{GonzalezDiaz:2007gt}, applied the method of Babichev  et al. to
a 
non-static generalisation of the Morris-Thorne metric, introducing a
shape function with an arbitrary dependence 
on time, $K(r,t)$. They recovered again Eq.~(\ref{once}) for the
temporal mass rate of a wormhole in the asymptotic 
limit. It must be noted that in their derivation of such expression
they allowed an arbitrary dependence of the 
energy density, pressure and the four velocity of the fluid on both
the time and radial coordinates. Therefore, the 
wormhole mass rate expression Eq.~(\ref{once}) must be valid in general
for asymptotically flat wormholes. It must be 
emphasised that a wormhole is a non-vacuum solution and that the
consideration of a time dependence in the shape 
function leads also to a non zero $\Theta_0^r$, which take into
account the non-zero energy flow onto the hole, therefore 
taking into account the back reaction.
Nevertheless, the authors of Ref.~\cite{GonzalezDiaz:2007gt} considered
that the most crucial argument against the big trip 
included in the paper of Faraoni \cite{Faraoni:2007kx} is the
vanishing of the proper radial velocity at the big rip and 
its quickly decrease close to it, since it is in the point of
introducing an explicitly equations of state for the fluid 
in order to integrate Eq.~(\ref{once}) where they were considering an
approximation. First of all, the authors noted 
that, besides the fact that the time where the universe is
engulfed by a wormhole is not only before the big rip 
but even before the divergence of the wormhole mouth, the important
quantity which refers to an accretion process is the 
proper radial flow, which is approximately $\rho v\sim a^{-3(1+w)/2}$
which increases with time for $w<-1$ and diverges 
at the big rip, what guarantees the process would not be stopped. In the
second place, they point out that the accretion of 
dark and phantom energy onto astronomical objects differs from the
accretion of usual energy concentrated in a given region 
of space onto those objects, because in the first case the
energy pervades the whole space being, therefore, a 
phenomenon not based on any fluid motion, but on increasing more and more space filled with such kind of energy inside the boundary of the considered object.

Later on, C.~Gao, X.~Chen, V.~Faraoni and Y.~G.~Shen,
\cite{Gao:2008jv}, emphasising that the method of Babichev 
et al. applied to black holes is not taking into account the
backreaction of the fluid on the background, claimed that the
results obtained by using that method can only be valid in a low
matter density background. In that spirit, they used a 
generalised McVittie metric and inserted a radial heat flux
term in the energy-momentum tensor, in order to show that 
a cosmological black hole (non-asymptotically flat) should increase by
the accretion of phantom energy. The authors also 
pointed out any difficulties to compare their solution with the
corresponding of Babichev 
et al. \cite{Babichev:2004yx}, arguing that this fact could be due to the simplifications taken in both cases.

The work of Gao et al. originated some interesting
comments. First, in Ref.~\cite{Zhang:2007yu}, X.~Zhang pointed 
out the shortcomings of the Babichev et al. method, in particular the
use of a non-cosmological metric and the exclusion 
of the backreaction of the phantom fluid on the black hole
metric. Nevertheless, Zhang found the results of 
Ref. \cite{Gao:2008jv} highly speculative, giving the example of the
use of a hypothesised metric. Therefore, he decided 
to use for the moment the method of Babichev et al. in order to
extract at least some tentative conclusions, lacking 
a complete method to the study of the dark energy accretion
phenomenon. Second, in \cite{MartinMoruno:2008vy} 
where a first attempt to include cosmological effects in the
study of the accretion of dark energy onto black 
holes was done by the consideration of an Schwarzschild-de Sitter
spacetime, it was noted that the results 
achieved in  \cite{Gao:2008jv}
were obtained under the assumption of a premise (contradictory with
all studies of this problem in the literature) in which 
their desired result is contained, making circular their whole
argument, and their result invalid. Third, in
\cite{Babichev:2008jb}, Babichev 
et al. included some comments expressing their doubts regarding the
conclusions of \cite{Gao:2008jv}. They claimed that 
the heat flux term is introduced in an unnatural way in the solution
of Gao et al. to support their configuration, because 
the perfect fluid is not accreted in the mentioned solution and that
such an introduction could may be lead to 
instabilities to small perturbations. They also pointed out that the temperature of the fluid blows up at the event horizon.

In summary, regarding the phenomenon of dark and phantom energy
accretion onto black holes the method developed by 
Babichev et al. \cite{Babichev:2004yx} has been improved to take into
account the backreaction on the black hole 
size \cite{MartinMoruno:2006mi} in an asymptotically flat space and
also in a cosmological one \cite{MartinMoruno:2008vy} 
by using a non-static generalisation of the Schwarzschild and
Schwarzschild-de Sitter metric, respectively. 
Nevertheless, the study still lacks the consideration of the
complete backreaction of the dark or phantom fluid in 
an asymptotically dark or phantom universe. 

Although the method used to study the dark and phantom energy
accretion onto wormholes is similar to the one treating 
the above mentioned phenomenon, in the case of the phantom energy
accretion onto wormholes the backreaction originated 
by the consideration of the phantom fluid is automatically taken into
account by using a non-static generalisation 
of the Morris-Thorne metric \cite{GonzalezDiaz:2007gt}. In this case
there are no studies, up to our knowledge, considering 
rigorously the wormhole accretion phenomenon in a cosmological spacetime.

We want however to point out, that the increase (decrease) of the
black hole size by the accretion of dark (phantom) energy, and the 
contrary in the case of wormholes, seems considerably  well supported. Such
affirmation can be also understood taken into account a 
different method. It is well known that the formalism developed by
Hayward for spherically symmetric spacetimes \cite{Hayward:2004fz}
implies 
that a dynamical black hole (characterised by a future outer trapping
horizon) would increase if it is considering in an 
environment fulfilling $p+\rho>0$ and decrease if $p+\rho<0$, phenomenon which is due to 
a flow of such surrounding material into the hole. Therefore, this
totally 
independent study confirms the results presented in this chapter about
black holes at least in a qualitative way. The 
question would be, how large are the quantitative differences which
could appear by the consideration of the 
backreaction and the cosmological space?

On the other hand, regarding wormholes it has been shown
\cite{MartinMoruno:2009iu} that in order to recover using 
the Hayward formalism the results obtained by the accretion method
where the backreaction is included and 
by the very basis of wormhole physics, a wormhole must be
characterised by a past outer trapping horizon. Since it seems 
that there would be no reason to change the local characterisation of
an astronomical object because of the consideration 
of such an object in a space with a different asymptotically
behaviour, the qualitative increase (decrease) of wormholes 
by the accretion of phantom (dark) energy should be recovered by
considering cosmological wormholes spacetimes. 
Nevertheless, whether a wormhole would suffer a so huge increase to
include the whole universe, occurring a 
big trip, is still an open question.

We want to point out that there are several interesting
opened questions concerning accretion onto black holes. More
improvement in
the accretion theory is needed to take into account the backreaction of the space-time and
study
the situation where a perfect fluid approximation is not valid, what would
clarify whether a black hole can become a naked singularity?
A more detailed study about the spin or charge super-radiance is also needed, in order to show whether these processes could concur in such a way that
finally the cosmic censorship conjecture would keep hold. Preliminary
results
indicate that this is the case, but the possibility of violation of the cosmic censorship conjecture produced by the
process of dark energy 
accretion onto a charged or rotating black holes is still open.

Finally, it must be worth noticing that recent papers, \cite{MersiniHoughton:2008aw,MersiniHoughton:2009dp}, consider the
Babichev et al. method to propose a new observational dark energy
test. The main idea is based on the black hole mass change induced by
the dark energy accretion process which, as we have shown in this
chapter, is proportional to $1+w$. Although a direct observation of
this change is beyond our current detection possibilities, because
the time scale to produce a 
change in the black hole mass measurable with our present devices
would be too long, it would cause observable modifications in the
orbital radius of the supermassive black hole binaries, since the
black hole binaries would either merge in a more accelerated way than
expected if $1+w>0$ or the merging would be progressively stopped if
the dominant energy 
condition is violated, being also possible that the binaries would rip apart in the second case\footnote{The interested reader can look up in
  \cite{MersiniHoughton:2008aw,MersiniHoughton:2009dp} for more
  details.}.
At this moment there are two candidates,
Galaxy 0402+379 and Radio Galaxy OJ287, for the observation of the mentioned phenomenon what could provide us with more information 
about the nature of dark energy, and probably more interesting
candidates can be expected in the future.

\section{Conclusions and further comments.}

In the present chapter we have shown that if the current accelerating
expansion of the universe is explained in the framework of General
Relativity, then the consideration of a dynamical dark energy fluid
would produce other effects besides the modelling of that
acceleration. In this sense, the consideration of dark energy would
not be simply the consideration of an ether covering 
the whole space, since footprints of such a fluid could appear
in our Universe by observing the evolution of black- and
worm-holes. Whether such effects are measurable in practice, is a
question related to the 
accuracy of the observational data.

The effects in question regarding the dynamical evolution of black-
and worm-holes, would be an additional increase (decrease) of the
black hole size in the case that the dark energy fulfils (violates)
the dominant 
energy condition, and the contrary in the wormhole case. Even more, these effects would produce changes in the orbital radius of black hole binaries which could be large enough to be detected in practice, helping us to get new constrains 
to dark energy equation of state parameter.

If one considers the used test fluid approximation to hold in phantom
models possessing a singularity at a finite time in the future, then
the black holes would tend to disappear at that singularity, which
would never be reached since a big trip phenomenon may take place
before. On the other hand, in dark energy models with $w>-1$ black
holes would not engulf the whole universe, since the current mass of a
hole able to exceed the universe size in a finite time should be so
huge that 
it would be bigger than the mass of the observable Universe.

Although, by the arguments presented in this chapter, the qualitative evolution of the considered astronomical objects by the accretion of dark energy seems to be a solid result, the quantitative dynamical behaviour could differ
from the mentioned results, since at the final step we are considering the approximation that the surrounding fluid is a cosmological one. In the black hole case this approximation is even stronger, since the non-static generalisation of the Schwarzschild metric, though taking into account the radial flow, lacks of a consideration of the backreaction.

Finally, whether or not the above features studied
in this chapter can be taken to imply that certain dark energy models
are  more consistent 
than others is a matter that
will depend on both the intrinsic consistency of the different models
and the current and future observational data.

\vspace{0.5cm}

\textbf{Acknowledgements.} We acknowledge La Casa de Arag\'on to
provide us a relaxing and inspiring place where this chapter was
designed and discussed. JAJM thanks the financial support provided
by Fundaci\'on Ram\'on Areces and the DGICYT 
Research Project MTM2008-03754. P.~M.~M.
gratefully acknowledges the financial support provided by the I3P framework
of CSIC and the European Social Fund and by
a Spanish MEC Research Project No.FIS2008-06332/FIS.
Special thanks to Jodie Holdway for English revision of this chapter.

\end{document}